\documentclass[multicol,showpacs,aps]{revtex4}

\usepackage{graphicx}
\usepackage{feynmf}
\usepackage{psfig}

\newcommand{\la}{\left\langle}
\newcommand{\ra}{\right\rangle}
\newcommand{\be}{\begin{equation}}
\newcommand{\ee}{\end{equation}}
\newcommand{\bea}{\begin{eqnarray}}
\newcommand{\eea}{\end{eqnarray}}
\newcommand{\ba}{\begin{array}}
\newcommand{\ea}{\end{array}}
\newcommand{\bi}{\begin{itemize}}
\newcommand{\ei}{\end{itemize}}
\newcommand{\piulug}{$\Pi^{u<}_{u>}$}
\newcommand{\piulbg}{$\Pi^{u<}_{b>}$}
\newcommand{\piulbl}{$\Pi^{u<}_{b<}$}
\newcommand{\piblbg}{$\Pi^{b<}_{b>}$}
\newcommand{\piblug}{$\Pi^{b<}_{u>}$}

\newcommand{\eqnKolmMHD}
{\be \label{eqn:KolmMHD}
E^{\pm}(k) =  K^{\pm} \frac{\left( \Pi^{\pm} \right)^{4/3}}
 		          {\left( \Pi^{\mp} \right)^{2/3}} k^{-5/3}
\ee}
\newcommand{\eqnMHDubx}
{\bea 
\frac{\partial {\bf u}}{\partial t} +
{\bf (u \cdot \nabla) u} & = & 
- \nabla p + {\bf (b \cdot \nabla) b} + \nu \nabla^2 {\bf u} 
		\label{eqn:udot_x} \\
\frac{\partial {\bf b}}{\partial t} +
{\bf (u \cdot \nabla) b} & = & 
- {\bf (b \cdot \nabla) u} + \eta \nabla^2 {\bf b}
		\label{eqn:bdot_x} \\
\nabla \cdot {\bf u} & = & 0 \\
\nabla \cdot {\bf b} & = & 0  \eea}
\newcommand{\eqnMHDzx}
{\bea 
\frac{\partial {\bf z^{\pm}}}{\partial t} +
{\bf (z^{\mp} \cdot \nabla) z^{\pm}} & = & 
- \nabla p  + \nu_+ \nabla^2 {\bf z^{\pm}}
+ \nu_- \nabla^2 {\bf z^{\mp}} \\
\nabla \cdot {\bf z^{\pm}} & = & 0  \eea}

\begin{document}
\title{Field theoretic calculation of energy cascade rates in 
nonhelical magnetohydrodynamic turbulence}
\author{Mahendra\ K.\ Verma  
\email{email: mkv@iitk.ac.in}}
\affiliation{Department of Physics, Indian Institute of Technology,
Kanpur  --  208016, INDIA}

\date{October 19, 2002}

\begin{abstract}
Energy cascade rates and Kolmogorov's constant for nonhelical steady
magnetohydrodynamic turbulence have been calculated by solving the
flux equations to the first order in perturbation.  For zero cross
helicity and space dimension $d=3$, magnetic energy cascades from
large length-scales to small length-scales (forward cascade).  In
addition, there are energy fluxes from large-scale magnetic field to
small-scale velocity field, large-scale velocity field to small-scale
magnetic field, and large-scale velocity field to large-scale magnetic
field.  Kolmogorov's constant for magnetohydrodynamics is
approximately equal to that for fluid turbulence ($\approx 1.6$) for
Alfv\'{e}n ratio $0.5 \le r_A \le \infty$.  For higher
space-dimensions, the energy fluxes are qualitatively similar, and
Kolmogorov's constant varies as $d^{1/3}$.  For the normalized cross
helicity $\sigma_c \rightarrow 1$, the cascade rates are proportional
to $(1-\sigma_c)/ (1+\sigma_c)$, and the Kolmogorov's constants vary
significantly with $\sigma_c$.
\end{abstract}

\vspace{1cm}
\pacs{PACS numbers: 47.27Gs, 52.35.Ra, 91.25.Cw}

\maketitle

\section{Introduction}
There are various phenomenological, numerical, and theoretical result
on magnetohydrodynamics (MHD) turbulence.  Kraichnan \cite{Krai:65}
and Irosnikov \cite{Iros} proposed a first phenomenology for
homogeneous and isotropic MHD which connects energy spectrum and
cascade rates.  In this phenomenology, the kinetic and magnetic energy
spectrum ($E^u(k)$ and $E^b(k)$ respectively) are proportional to
$k^{-3/2}$, and the energy cascade rate of flux $\Pi = (E(k))^2
k^3/B_0$, where $B_0$ is the mean magnetic field.  In an alternate
phenomenology, Marsch \cite{Mars:Kolm}, Matthaeus and Zhou
\cite{MattZhou}, and Zhou and Matthaeus \cite{ZhouMatt} predicted that
energy spectrum is Kolmogorov-like ($E(k) \propto k^{-5/3}$) and $\Pi
= (E(k))^{3/2} k^{5/2}$.  Verma et al.~\cite{MKV:MHDsimu} numerically
calculated the energy cascade rates of Els\"{a}sser variable ${\bf u
\pm b}$ ({\bf u,b} are velocity and magnetic field fluctuations), and
found their results to be consistent with Kolmogorov-like
phenomenology, rather that that of Kraichnan and Irosnikov.  Frick and
Sokoloff \cite{Fric} studied the spectra and cascade rates in a shell
model of MHD turbulence; they found the spectral index to be close to
5/3 in absence of cross helicity and magnetic helicity.  According to
their study, the helicities suppress the cascade process.  Recently,
M\"{u}ller and Biskamp \cite{Bisk1:Kolm} and Biskamp and M\"{u}ller
\cite{Bisk2:Kolm} computed the spectral index and found it to be
closer to 5/3.  Kolmogorov-like spectrum is also supported by recent
theoretical results \cite{MKV:B0RG,Srid1,Srid2,MKV:MHD_PRE,MKV:MHDRG,Hnat}.

The energy cascade rates of MHD depend on cross helicity ($H_c = {\bf
u \cdot b}$), magnetic helicity ($H_M = {\bf a \cdot b}$, where 
${\bf a}$ is vector potential), and kinetic helicity ($H_K= {\bf u
\cdot \omega}$, where ${\bf \omega}$ is vorticity).  Pouquet {\em et
al.}~\cite{Pouq:EDQNM} applied EDQNM approximation to study  energy
fluxes.  For nonhelical MHD, Pouquet et al.~\cite{Pouq:EDQNM} argued
that the ME cascade is forward, i.e., from large-scale to small-scale.
However, in the presence of helicity, they observed that the
large-scale magnetic energy brings to equipartition the small-scale
kinetic and magnetic excitation by Alfv\'{e}n effect, and the
``residual helicity'', $H_K-H_M$, induces growth of large-scale
magnetic energy and helicity.  Pouquet and Patterson \cite{Pouq:num}
studied this problem using direct numerical simulation and arrived at
similar conclusions.  In the present paper we will derive the energy
fluxes of nonhelical MHD using field theoretic methods.  The energy
fluxes of helical MHD are discussed in a companion paper, Verma
\cite{MKV:MHDhelical}, referred to as paper II.

Since there are two fields {\bf u} and {\bf b} in MHD, the energy can
be transferred from {\bf u} to {\bf u}, {\bf u} to {\bf b}, and {\bf
b} to {\bf b}.  The resulting energy fluxes due to these transfers are
illustrated in Fig.~1.  These fluxes have been numerically calculated
recently by Dar {\em et al.}~\cite{Dar:flux} and Ishizawa and Hattori
\cite{Ishi:flux} for two-dimensional (2D) MHD turbulence.  The
numerical values calculated by Dar {\em et al.}  for $\sigma_c \approx
0$ and $r_A$ (KE/ME) $\approx 0.5$ are listed in
Table~\ref{tab:flux_Dar}.  The prime conclusions of Dar {\em et
al.}~\cite{Dar:flux} and Ishizawa and Hattori's \cite{Ishi:flux} 2D
numerical study are: (1) the ME cascades from large-scales to
small-scales (forward cascade); (2) there is a significant energy
transfer from large length-scale velocity field to large length-scale
magnetic field; this transfer could play an important role in ME
enhancement; (3) there is an inverse cascade of KE.  In another recent
work, Cho and Vishniac \cite{ChoVish:gene} have derived some
interesting scaling relationships between the energy transfer rates
and verified them using numerical simulations.  Since the direction of
total energy cascade is the same in 2D and 3D (three-dimensional) MHD
turbulence, some of the conclusions drawn by Dar {\em et
al.}~\cite{Dar:flux} and Ishizawa and Hattori \cite{Ishi:flux} based
on 2D simulations are expected to hold at least qualitatively in 3D
MHD turbulence.  Therefore, in this paper we compare the
numerical results of Dar et al.~\cite{Dar:flux} and Ishizawa and
Hattori \cite{Ishi:flux} with our 3D analytic results.

In the present paper we have carried out the energy cascade rate
calculation for MHD turbulence for the {\em inertial-range
wavenumbers} using perturbative field-theoretic technique.  Here we
assume that the turbulence is homogeneous and isotropic to make the
problem tractable.  Even though the real-world turbulence does not
satisfy these properties, at least at the large-scales, many
conclusions drawn using this assumption give us important insights
into the energy transfer mechanisms, as will be discussed in this
paper and paper II.  We assume that the mean magnetic field is absent;
this assumption is to ensure that the turbulence is isotropic.  Our
procedure requires Fourier space integrations of functions involving
products of energy spectrum and the Greens functions.  Since there is
a general agreement on Kolmogorov-like spectrum for MHD turbulence, we
take $E(k) \propto k^{-5/3}$ for all the energy spectra for MHD.  For
the Greens function, we substitute the ``renormalized'' or ``dressed''
Greens function computed by Verma \cite{MKV:MHDRG}.  After this
substitution, various energy fluxes and Kolmogorov's constant of MHD
are computed. Using the steady-state condition, we also calculate the
energy supply from the large-scale velocity field to the large-scale
magnetic field.  This result is quite robust, and is independent of
the nature of large-scale forcing.  In this paper we assume both
kinetic and magnetic helicity to be absent.  The energy fluxes for
helical MHD are discussed in paper II.

The parameter space of MHD is rather large because of various energy
spectra.  The two well known parameters are: the normalized cross
helicity $\sigma_c$, which is the ratio of twice cross helicity and
energy; and the Alfv\'{e}n ratio $r_A$, which is the ratio of kinetic
energy and magnetic energy.  Calculation of cascade rates for
arbitrary $\sigma_c$ and $r_A$ is quite complex.  In this paper we
limit ourselves to two limiting cases: (1) $\sigma_c=0$ and whole
range of $r_A$; (2) $\sigma_c \rightarrow 1$ and $r_A=1$.  Strictly
speaking, the parameters used in our calculations are spectral
$\sigma_c(k)= 2 H_c(k)/(E^u(k)+E^b(k))$ and $r_A(k)=E^u(k)/E^b(k)$.
Since our calculation is confined to inertial-range wavenumbers where
Kolmogorov's 5/3 powerlaw is valid for all the energy spectra of MHD,
both $\sigma(k)$ and $r_A(k)$ can be treated as constants.  Note that
these parameters may differ from the global $\sigma_c$ and $r_A$. We
carry out our theoretical analysis in various dimensions.  We will
show that our theoretical results are in general agreement with the
simulation results of Dar {\em et al.}~\cite{Dar:flux} and Ishizawa
and Hattori \cite{Ishi:flux}.

The outline of this paper is as follows: In section 2 we calculate
various cascade rates for $\sigma_c=0$ case.  The other extreme case
$\sigma_c \rightarrow 1$ is considered in section 3.  Section 4
contains summary and conclusions.

\section{Cascade Rates in MHD Turbulence: $\sigma_{c} =0$}

In this section we will analytically compute the energy cascade rates
when $\sigma_c=0$.  We take the following form of Kolmogorov's
spectrum for kinetic energy (KE) and magnetic energy (ME)
\bea 
E^u(k) & = & K^u \Pi^{2/3} k^{-5/3} \label{eqn:Euk} \\
E^b(k) & = & E^u(k) / r_A  \label{eqn:Ebk}
\eea
where $K^u$ is Kolmogorov's constant for MHD turbulence, and $\Pi$ is
the total energy flux.  Another Kolmogorov's constant $K$ is defined
for the total energy,
\be
E_{total}(k) = E^u(k)+E^b(k) = E(k) = K \Pi^{2/3} k^{-5/3},
\ee
with 
\be
K = K^u (1+r_A^{-1})  \label{eqn:Kplus}
\ee
With this preliminaries we start our flux  calculation.

The incompressible MHD equations are
\eqnMHDubx
where ${\bf u}$ and ${\bf b}$ are the velocity and magnetic fields
respectively, $p$ is the total pressure, and $\nu$ and $\mu$ are the
kinematic viscosity and magnetic diffusivity respectively.  To compute
various energy transfers among various Fourier modes we resort to the
energy equations, which are \cite{Pouq:EDQNM,Dar:flux,Stan:book}
\bea
\left(\frac{\partial}{\partial t}  + 2 \nu k^2 \right) C^{uu}({\bf k},t,t) 
& = &	\frac{1}{(d-1)(2 \pi)^d} 
	\int_{\bf k'+p+q=0} \frac{d {\bf p}}{(2 \pi)^d} 
	       [S^{uu}({\bf k'|p|q})+S^{uu}({\bf k'|q|p}) \nonumber \\
&   & 	\hspace{2in}+S^{ub}({\bf k'|p|q})+S^{ub}({\bf k'|q|p})] \\
\left(\frac{\partial}{\partial t}  + 2 \eta k^2 \right) C^{bb}({\bf k},t,t) 
& = &	\frac{1}{(d-1)(2 \pi)^d} 
	\int_{\bf k'+p+q=0} \frac{d {\bf p}}{(2 \pi)^d} 
	       [S^{bu}({\bf k'|p|q})+S^{bu}({\bf k'|q|p}) \nonumber \\
&   & 	\hspace{2in}+S^{bb}({\bf k'|p|q})+S^{bb}({\bf k'|q|p})] 
\eea
The above
integrals have constraints that ${\bf k'+p+q = 0}$ (${\bf k=-k'}$).
The equal-time correlation functions used in the
energy equations are defined using
\bea
\left\langle u_i ({\bf p},t) u_j ({\bf q},t)\right\rangle  & = &
P_{ij}({\bf p)} C^{uu} ({\bf p},t,t) \delta({\bf p+q}) 
	\label{eqn:uiuj} \\
\left\langle b_i ({\bf p},t) b_j ({\bf q},t) \right\rangle  &= &
P_{ij}({\bf p)} C^{bb}({\bf p},t,t) \delta({\bf p+q}) 
	\label{eqn:bibj} \\
\left\langle u_i ({\bf p},t) b_j ({\bf q},t)\right\rangle  &= &
P_{ij}({\bf p)} C^{ub} ({\bf p},t,t) \delta({\bf p+q}) \label{eqn:avgend}
\eea
and the energy transfer rates $S({\bf k'|p|q})$ are defined using
\bea
S^{uu}({\bf k'|p|q}) & = &
   -\Im ({\bf [k' \cdot u(q)][u(k') \cdot u(p)]}),
	\label{eqn:Suu} \\
S^{bb}({\bf k'|p|q}) & = &
   -\Im ({\bf [k' \cdot u(q)][b(k') \cdot b(p)]}), 
	\label{eqn:Sbb} \\
S^{ub}({\bf k'|p|q}) & = &
    \Im ({\bf [k' \cdot b(q)][u(k') \cdot b(p)]}), 
	\label{eqn:Sub}  \\
S^{bu}({\bf k'|p|q}) & = & - S^{ub}({\bf p|k'|q})  
	\label{eqn:Sbu}
\eea
Here $\Im$ stands for the imaginary part of the argument.  
Note that $C^{ub}=0$ because $\sigma_c$ has been taken to be zero.
The above equations are based on Dar et al.'s formalism, which is a
generalization of those of Pouquet {\em et al.}~\cite{Pouq:EDQNM} and
Stani\u{s}i\'{c} \cite{Stan:book} and others.  In Dar {\em et al.}'s
formalism, the terms $S({\bf k|p|q})$ represents energy transfer from
mode ${\bf p}$ (the second argument of $S$) to ${\bf k}$ (the first
argument of $S$) with mode ${\bf q}$ (the third argument of $S$)
acting as a mediator.  Note that in the expression for $S$, the field
variables with the first and second arguments are dotted together,
while the field variables with the third argument is dotted with the
wavevector ${\bf k}$.  Dar {\em et al.}'s formulas has certain
advantages over those of Pouquet {\em et al.}~\cite{Pouq:EDQNM} and
Stani\u{s}i\'{c} \cite{Stan:book}.  Some of the quantities to be
defined below were not accessible in the earlier formalism, but now
they can be calculated using Dar {\em et al.}'s formulas.  In
addition, the flux formulas derived using the new scheme are
relatively simpler.

After some algebraic manipulation it can be shown that
\bea
S^{uu}({\bf k'|p|q}) + S^{uu}({\bf k'|q|p}) + 
S^{uu}({\bf p|k'|q}) + S^{uu}({\bf p|q|k'}) + 
S^{uu}({\bf q|k'|p}) + S^{uu}({\bf q|p|k'}) & = & 0  \\
S^{bb}({\bf k'|p|q}) + S^{bb}({\bf k'|q|p}) + 
S^{bb}({\bf p|k'|q}) + S^{bb}({\bf p|q|k'}) + 
S^{bb}({\bf q|k'|p}) + S^{bb}({\bf q|p|k'}) & = & 0  \\
S^{ub}({\bf k'|p|q}) + S^{ub}({\bf k'|q|p}) + 
S^{ub}({\bf p|k'|q}) + S^{ub}({\bf p|q|k'}) + 
S^{ub}({\bf q|k'|p}) + S^{ub}({\bf q|p|k'})  \nonumber \\
+S^{bu}({\bf k'|p|q}) + S^{bu}({\bf k'|q|p}) + 
S^{bu}({\bf p|k'|q}) + S^{bu}({\bf p|q|k'}) + 
S^{bu}({\bf q|k'|p}) +S^{bu}({\bf q|p|k'}) & = & 0 
\eea
These are the statements of  ``detailed conservation of energy''
in MHD triads  (when $\nu=\eta=0$) \cite{Lesl:book}.

For energy flux study, we split the wavenumber space into two regions:
$k<k_0$ (inside ``$k_0$-sphere'') and $k>k_0$ (outside
``$k_0$-sphere'').  This division is done for both velocity and
magnetic fields.  The energy transfer could take place from
inside/outside u/b-sphere to inside/outside u/b-sphere.  In terms of
$S$, the energy flux from inside of the $X$-sphere to outside of the
$Y$-sphere is
\bea 
\Pi^{X<}_{Y>}(k_0) & = & \int_{k'>k_0} \frac{d {\bf k}}{(2 \pi)^d} 
		       \int_{p<k_0} \frac{d {\bf p}}{(2 \pi)^d}  
			\la S^{YX}({\bf k'|p|q}) \ra
\label{eqn:flux}		
\eea 
where $X$ and $Y$ stand for $u$ or $b$.  The energy fluxes from inside
$u/b$-sphere to outside $u/b$-sphere can be calculated by earlier
formalism, as well as that of Dar {\em et al.}  However, the fluxes from
inside $u$-sphere to inside $b$-sphere, and outside $u$-sphere to
outside $b$-sphere can be numerically calculated only by Dar {\em et al.}'s
formalism.  In this paper we will analytically calculate the above
fluxes in the inertial range using the Kolmogorov-like energy spectrum.

We assume that the kinetic energy is forced at small wavenumbers, 
and the turbulence is steady.  Therefore,
\bea
\Pi^{u<}_{b<} & = & \Pi^{b<}_{b>} + \Pi^{b<}_{u>} \label{eqn:ulbl} \\
\mbox{Input Kinetic Energy} & = & \Pi^{u<}_{u>} + \Pi^{u<}_{b>} 
				+ \Pi^{u<}_{b<}
\eea
We calculate the energy flux
$\Pi^{u<}_{b<}$ using the above steady-state property.  Hence the energy
feed into the large-scale magnetic field from the large-scale
velocity field could be obtained theoretically irrespective of
nature of large-scale forcing.  

We will analytically calculate the above energy  fluxes [Eq. (\ref{eqn:flux})]
in the inertial range to the leading order in perturbation series.  It
is assumed that ${\bf u(k)}$ is gaussian to leading order. Consequently, the
ensemble average of $S^{YX}$, $\la S^{YX} \ra$, is zero to the zeroth
order, but is nonzero to the first order.  The first order terms for
$S^{YX}(k|p|q)$ in terms of Feynman diagrams are shown below:
\input{flux_feyn.diag}

In the above diagrams the solid, dashed, wiggly (photon), and curly
(gluons) lines denote $\la u_i u_j \ra, \la b_i b_j \ra, G^{uu}$, and
$G^{bb}$ respectively.  In all the diagrams, the left vertex denotes
$k_i$, while the filled circle and the empty circles of right vertex
represent $(-i/2) P^+_{ijm}$ and $-i P^-_{ijm}$ respectively. Since
$G^{ub},G^{bu},C^{ub}$, and $C^{bu}$ are zero when $\sigma_c=0$, we
have not included the Feynman diagrams containing these terms.
 When we
substitute $\la u_i u_j \ra, \la b_i b_j \ra$ using
Eqs.~(\ref{eqn:uiuj},\ref{eqn:bibj}), we obtain terms involving
$C^{X}(p,t,t') C^{Y}(q,t,t')$.
The resulting expressions for various $\la S^{YX}(k|p|q) \ra$ are
\bea
\la S^{uu}(k|p|q)\ra & = &\int_{-\infty}^t  dt'  [
                T_{1}(k,p,q) G^{uu}(k,t-t') C^{uu}(p,t,t') C^{uu}(q,t,t') 
			\nonumber \\
& &\hspace{1cm} + T_{5}(k,p,q) G^{uu}(p,t-t') C^{uu}(k,t,t') C^{uu}(q,t,t') 
			\nonumber \\
& &\hspace{1cm} + T_{9}(k,p,q) G^{uu}(q,t-t') C^{uu}(k,t,t') C^{uu}(p,t,t') ]
			\label{eqn:flux_uu}	\\
\la S^{ub}(k|p|q) \ra & =- & \int_{-\infty}^t  dt'  [
                T_{2}(k,p,q) G^{uu}(k,t-t') C^{bb}(p,t,t') C^{bb}(q,t,t')
			\nonumber \\
& &\hspace{1cm} + T_{7}(k,p,q) G^{bb}(p,t-t') C^{uu}(k,t,t') C^{bb}(q,t,t')
			\nonumber \\
& &\hspace{1cm} + T_{11}(k,p,q) G^{uu}(q,t-t') C^{uu}(k,t,t') C^{bb}(p,t,t') ]
			\\
\la S^{bu}(k|p|q) \ra & = & -\int_{-\infty}^t  dt'  [
                T_{3}(k,p,q) G^{bb}(k,t-t') C^{uu}(p,t,t') C^{bb}(q,t,t')
			\nonumber \\
& &\hspace{1cm} + T_{6}(k,p,q) G^{uu}(p,t-t') C^{bb}(k,t,t') C^{bb}(q,t,t')
			\nonumber \\
& &\hspace{1cm} + T_{12}(k,p,q) G^{bb}(q,t-t') C^{bb}(k,t,t') C^{uu}(p,t,t') ]
			\\
\la S^{bb}(k|p|q) \ra & = & \int_{-\infty}^t  dt'  [
                T_{4}(k,p,q) G^{bb}(k,t-t') C^{bb}(p,t,t') C^{uu}(q,t,t')
			\nonumber \\
& &\hspace{1cm} + T_{8}(k,p,q) G^{bb}(p,t-t') C^{bb}(k,t,t') C^{uu}(q,t,t')
			\nonumber \\
& &\hspace{1cm} + T_{10}(k,p,q) G^{uu}(q,t-t') C^{bb}(k,t,t') C^{bb}(p,t,t') ]
\eea
where $T_i (k,p,q)$ are functions of wavevectors $k,p$, and $q$ given in 
Appendix A.

The Greens functions can be written in terms of ``effective'' or
``renormalized'' viscosity $\nu(k)$ and resistivity $\eta(k)$ (see
Verma \cite{MKV:MHDRG} for details) as
\bea
G^{uu}(k,t-t') & = & \exp\left(- \nu(k) k^2 (t-t') \right) \\ 
G^{bb}(k,t-t') & = & \exp\left(- \eta(k) k^2 (t-t') \right) 
\eea
The relaxation time for $C^{uu}(k,t,t')$ is assumed to be the same as
that of $G^{uu}$, and that of $C^{bb}(k,t,t')$ is assumed to be the same
as that of $G^{bb}$.  Therefore the time dependence of the
unequal-time correlation functions will be
\bea
C^{uu}(k,t,t') & = & \exp \left(- \nu(k) k^2 (t-t') \right) 
			C^{uu}(k,t,t) \\ 
C^{bb}(k,t,t') & = & \exp \left(- \eta(k) k^2 (t-t') \right) 
			C^{bb}(k,t,t) 
\eea
The above forms of Green's and correlation functions are substituted in
the expression of $\la S^{YX} \ra$, and the $t'$ integral is
performed.  Now Eqs.~(\ref{eqn:flux}, \ref{eqn:flux_uu}) yield the
following flux formula for $\Pi^{u<}_{u>}(k_0)$:
\bea
\Pi^{u<}_{u>}(k_0) & = & \int_{k>k_0} \frac{d {\bf k}}{(2 \pi)^d} \int_{p<k_0} 
                         \frac{d {\bf p}}{(2 \pi)^d}  
   	                 \frac{1}{\nu(k) k^{2}+ 
				\nu(p) p^{2}+\nu(q) q^{2}}
			\times \nonumber \\
& &  [ T_{1}(k,p,q) C^{uu}(p) C^{uu}(q)
    +T_{5}(k,p,q) C^{uu}(k) C^{uu}(q)
     +T_{9}(k,p,q) C^{uu}(k) C^{uu}(p) ] 
    \label{eqn:Pi_mhd}
\eea
The expressions for the other fluxes can be obtained similarly.

The equal-time correlation functions $C^{uu}(k,t,t)$ and $C^{bb}(k,t,t)$
at the steady-state can be written in terms of one dimensional energy spectrum as
\begin{eqnarray}
C^{uu}(k,t,t) = \frac{2 (2 \pi)^d}{S_d (d-1)} k^{-(d-1)} E^{u}(k) \\
C^{bb}(k,t,t) = \frac{2 (2 \pi)^d}{S_d (d-1)} k^{-(d-1)} E^{b}(k)
\end{eqnarray}
where $S_d$ is the surface area of $d$-dimensional unit spheres. 
We are interested in the fluxes in the inertial range.  Therefore, we
substitute Kolmogorov's spectrum [Eqs.~(\ref{eqn:Euk},\ref{eqn:Ebk})]
for the energy spectrum.  The effective viscosity and resistivity are
proportional to $k^{-4/3}$, i.e.,
\bea
\nu(k) & = & (K^u)^{1/2} \Pi^{1/3} k^{-4/3} \nu^*    
	 \label{eqn:nuk} \\
\eta(k) & = & (K^u)^{1/2} \Pi^{1/3} k^{-4/3} \eta^*  .
	\label{eqn:etak}
\eea
and the parameters $\nu^*$ and $\eta^*$ were calculated in Verma 
\cite{MKV:MHDRG}.  

The $d$-dimensional volume integral under the constraint ${\bf
k'+p+q=0}$ is given by \cite{FourFris}
\be
\int_{\bf p+q=k} d{\bf q}  = S_{d-1} \int dp dq 
	\left( \frac{p q}{k} \right)^{d-2} (\sin \alpha)^{d-3}
\ee
where $\alpha$ is angle between vectors ${\bf p}$ and ${\bf q}$.  We
also nondimensionalize Eq.~(\ref{eqn:Pi_mhd}) by substituting
\cite{Lesl:book}
\be
k=\frac{k_0}{u}; \hspace{1cm} p=\frac{k_0}{u} v; 
	\hspace{1cm} q=\frac{k_0}{u} w
\ee
which yields
\be
\Pi^{X<}_{Y>} =  (K^u)^{3/2} \Pi \left[ \frac{4 S_{d-1}}{(d-1)^2 S_d}
              \int_0^1 dv \ln{(1/v)} \int_{1-v}^{1+v} dw 
		(vw)^{d-2} (\sin \alpha)^{d-3}
		F^{X<}_{Y>}(v,w) \right]
\label{eqn:piub}
\ee
where the integrals $F^{X<}_{Y>}(v,w)$ are
\bea
F_{u>}^{u<} & = &  \frac{1}{\nu^*(1+v^{2/3}+w^{2/3})} 
		[t_{1}(v,w) (v w)^{-d-\frac{2}{3}}
		+t_{5}(v,w) w^{-d-\frac{2}{3}} 
                +t_{9}(v,w) v^{-d-\frac{2}{3}} ] 
		\label{eqn:Fuu} \\
F_{u>}^{b<} & = &  \frac{-1}{\nu^*+\eta^*(v^{2/3}+w^{2/3})} 
		[ t_{2}(v,w) (v w)^{-d-\frac{2}{3}} r_A^{-2}
		+t_{7}(v,w) w^{-d-\frac{2}{3}} r_A^{-1} 
		+t_{11}(v,w) v^{-d-\frac{2}{3}}  r_A^{-1}] 
			\label{eqn:Fub} \\
F_{b>}^{u<} & = &  \frac{-1}{\nu^* v^{2/3}+\eta^*(1+w^{2/3})} 
		[t_{3}(v,w) (v w)^{-d-\frac{2}{3}} r_A^{-1}
		+t_{6}(v,w) w^{-d-\frac{2}{3}} r_A^{-2} 
		+t_{12}(v,w)v^{-d-\frac{2}{3}}  r_A^{-1}] 
			\label{eqn:Fbu} \\
F_{b>}^{b<} & = &  \frac{1}{\nu^* w^{2/3}+\eta^*(1+v^{2/3})} 
		[t_{4}(v,w) (v w)^{-d-\frac{2}{3}} r_A^{-1}
		+t_{8}(v,w) w^{-d-\frac{2}{3}} r_A^{-1} 
		+t_{10}(v,w) v^{-d-\frac{2}{3}}  r_A^{-2}]  
			\label{eqn:Fbb}.
\eea
Here $t_i(v,w)=T_i(k,kv,kw)/k^2$.
Note that the energy fluxes are constant consistent with the Kolmogorov's
picture. We compute the bracketed terms (denoted by $I^{X<}_{Y>}$) 
numerically and find that all of them converge. 
Let us denote $I = I_{u>}^{u<} + I_{u>}^{b<} + I_{b>}^{u<} + I_{b>}^{b<}$.
Using the fact that  the total flux $\Pi$ is 
\be
\Pi = \Pi_{u>}^{u<} + \Pi_{u>}^{b<} + \Pi_{b>}^{u<} + \Pi_{b>}^{b<},
\ee
we can calculate the value of constant $K^u$, which is 
\be
K^u = (I)^{-2/3} 
\ee
In addition, the energy flux ratios can be computed using $\Pi^{X<}_{Y>}/\Pi
= I^{X<}_{Y>}/I$.  The flux ratio  $\Pi^{u<}_{b<}/\Pi$
is obtained using steady-state condition [Eq.~(\ref{eqn:ulbl})].
The values of constant $K$ can be computed using
Eq.~(\ref{eqn:Kplus}). The flux ratios and Kolmogorov's constants for
$d=3$ and various $r_A$ are listed in Table~\ref{tab:flux_d3}.  The
same quantities for $r_A=1$ and various space dimensions are listed in
Table~\ref{tab:flux_alldra1}.

The following trends can be inferred by studying
Table~\ref{tab:flux_d3}.  We find that for $d=3$, \piulug/$\Pi$ starts
from 1 for large $r_A$ and decreases nearly to zero near $r_A=0.3$.
All other fluxes start from zero and increase up to some saturated
values; this implies that near $r_A \approx 1$, all the energy  fluxes become
significant.  Clearly, the sign of \piblbg is positive, indicating
that ME cascades from large length-scale to small length-scale.  Under
steady-state the large-scale ME is maintained by \piulbl, which is one
of the most dominant transfers near $r_A = 0.5$.  The energy
flux $\Pi^{u<}_{b<}$ entering the large scales magnetic energy could
play an important role in the amplification of magnetic energy.  In
paper II we will construct a dynamo model based on the  energy fluxes.

The Kolmogorov constant $K$ for $d=3$ is listed in
Table~\ref{tab:flux_d3}.  For all $r_A$ greater than 0.5, $K$ is
approximately constant and is close to 1.6, same as that for fluid
turbulence ($r_A=\infty$).  Since $\Pi \propto K^{-3/2}$, we can
conclude that the variation of $r_A$ (redistribution of fluid and
magnetic energy, keeping the total energy fixed) does not change the
total cascade rate.  Near $r_A=0.3$, the constant $K$ appears to
increase, indicating a sudden drop in the cascade rate.  When $r_A$ is
decreased further, near $r_A=0.25$ both $\nu^* \approx 0$ and
$\eta^* \approx 0$ \cite{MKV:MHDRG}, or $K \rightarrow \infty$.
This signals an absence of turbulence for $r_A$ near 0.25.  This is
consistent with the fact that MHD equations are linear in the $r_A
\rightarrow 0$ (fully magnetic) limit, hence do not exhibit
turbulence.  However, it is still surprising that turbulence
disappears near $r_A=0.25$ itself.

The Kolmogorov's constant $K$ computed above can be used to estimate
the amount of turbulent heating in the solar wind.  Verma et
al.~\cite{MKV:SWheat} and Tu \cite{Tu} have put constraints on the
turbulent heating in the solar wind from the radial variation of
temperature in the solar wind.  Verma {\em et al.} \cite{MKV:SWheat}
observed that when $K \approx 1$, all the heating in the solar wind
for streams with $\sigma_c \approx 0$ can be accounted for by the
turbulent heating.  Our theoretical value for this constant in the
absence of mean magnetic field is approximately 1.5, larger than 1.
If we take $K \approx 1.5$ for solar wind streams with $\sigma_c
\approx 0$, only a fraction, possibly around half ($(1/1.5)^{3/2}$),
of the heating will be due to turbulence.  However, neglect of mean
magnetic field, anisotropy, helicities etc. are gross assumptions, and
we can only claim general consistency of the theoretical estimates
with the observational results of the solar wind.

We have calculated the flux ratios and the constant $K$ for various
space dimensions $d \ge 2.2$.  In Verma \cite{MKV:MHDRG} it has been
shown that for $d < 2.2$, the RG fixed point is unstable, and the
renormalized parameters could not be determined.  Due to that reason
we have calculated fluxes and Kolmogorov's constant for $d \ge 2$
only.  For these calculations we take $r_A=1$, which is a generic
case.  The calculated values are shown in
Table~\ref{tab:flux_alldra1}.  It is striking that all the fluxes are
approximately the same for large $d$.  In addition, \piulbl/$\Pi$ is
approximately 0.5 for all dimensions greater than 4.

We verify that $I^{X<}_{Y>}$ for constant $\nu^*$ and $\eta^*$ are
proportional to $d^{-1}$.  In Verma \cite{MKV:MHDRG} we
find that $\nu^*, \eta^* \propto d^{-1/2}$.  Therefore, $K
\propto d^{-1/3}$.
This result is a generalization of theoretical analysis of Fournier
{\em et al.} \cite{FourFris} for fluid turbulence.

In this section we calculated the cascade rates for $\sigma_c=0$.
In the next section we take the other limit 
$\sigma_c \rightarrow 1$.

\section{Cascade Rates in  MHD Turbulence: $\sigma_c \rightarrow 1$}

In this section we will describe the calculation of the energy cascade
rates for the large normalized cross-helicity ($\sigma_c \rightarrow
1$), and show that the cascade rates crucially depend on cross
helicity.  For cases with $\sigma_c \rightarrow 1$, it is best to work
with Els\"{a}sser variables ${\bf z^{\pm} = u \pm b}$.  For the
following discussion we will denote the ratio $\la
|z^-|^2\ra/\la|z^+|^2\ra$ by $r$.  Clearly $r \ll 1$.  Here we limit
ourselves to $r_A=1$.

The incompressible MHD equations in terms of ${\bf z^{\pm}}$  are
\eqnMHDzx
where $p$ is the total pressure, and $\nu_{\pm} = (\nu \pm \eta)/2$. 
Numerical simulations of Verma et al.~\cite{MKV:MHDflux} and
Dar \cite{Dar:thesis}, solar wind observations of Matthaeus and
Goldstein \cite{MattGold}, and  
Marsch and Tu \cite{MarsTu90}, and theoretical
calculations of Verma \cite{MKV:MHD_PRE,MKV:MHDRG} show that
Kolmogorov-like energy spectrum is valid even for
nonzero cross helicity, i.e.,
\eqnKolmMHD
where $K^{\pm}$ are Kolmogorov's constants for MHD.
The above equation was first derived by Marsch \cite{Mars:Kolm}.

The corresponding equations for the energy evolution are
\bea
\left(\frac{\partial}{\partial t}  + 2 \nu_+ k^2 \right)
C^{\pm \pm}({\bf k},t,t) + 2 \nu_- k^2 
C^{\pm \mp}({\bf k},t,t) 
& = &	\frac{1}{(d-1)(2 \pi)^d} 
	\int_{\bf k'+p+q=0} \frac{d {\bf p}}{(2 \pi)^d}  \nonumber \\
& &   [S^{\pm \pm}({\bf k'|p|q})+S^{\pm \pm}({\bf k'|q|p})] 
\eea
where
\be
S^{\pm \pm}({\bf k'|p|q}) = -\Im ({\bf [k' \cdot z^{\mp}(q)]
			[z^{\pm}(k') \cdot z^{\pm}(p)]}) 
\label{eqn:Szz}
\ee
and the equal-time correlations functions $C^{\pm \pm}$ and $C^{\pm \mp}$
are defined using
\bea
\left\langle z^{\pm}_i ({\bf p},t) z^{\pm}_j ({\bf q},t) 
	\right\rangle  & = &
P_{ij}({\bf p)} C^{\pm \pm} ({\bf p},t,t) \delta({\bf p+q}) \\
\left\langle z^{\pm}_i ({\bf p},t) z^{\mp}_j ({\bf q},t) 
	\right\rangle  & = &
P_{ij}({\bf p)} C^{\pm \mp}({\bf p},t,t) \delta({\bf p+q}) 
\eea
From Eq.~(\ref{eqn:Szz}) it is
evident that in the nonlinear transfers, the modes $z^+$ transfer
energy only to $z^+$ while $z^-$ acts as a mediator.  Similarly $z^-$
transfers energy only to $z^-$ with $z^+$ acting as a mediator.
It can be easily shown that
\be
S^{\pm}({\bf k'|p|q}) + S^{\pm}({\bf k'|q|p}) + 
S^{\pm}({\bf p|k'|q}) + S^{\pm}({\bf p|q|k'}) + 
S^{\pm}({\bf q|k'|p}) + S^{\pm}({\bf q|p|k'})  =  0 
\ee
These equations correspond to the ``detailed conservation of
energy'' in MHD triads.

In terms of $z^{\pm}$ variables, there are only two types of
fluxes $\Pi^{\pm}$, one for the $z^+$ cascade and the other for $z^-$
cascade.  In terms of $S$, these energy fluxes $\Pi^{\pm}$ are 
\bea 
\Pi^{\pm}(k_0) & = & \int_{k'>k_0} \frac{d {\bf k}}{(2 \pi)^d} 
		       \int_{p<k_0} \frac{d {\bf p}}{(2 \pi)^d}  
			\la S^{\pm \pm}({\bf k'|p|q}) \ra
\label{eqn:flux_zpm}		
\eea 
As described in the last section, the above fluxes are calculated
to the leading order in perturbation series.  To the  first order,
$\la S^{\pm \pm}({\bf k'|p|q}) \ra$ are
\bea
\label{eqn:Spm}
\la S^{\pm \pm}(k|p|q)\ra & = &\int_{-\infty}^t  dt'  [
		   T_{13}(k,p,q) G^{\pm \pm}(k,t-t') 
		   C^{\pm \mp}(p,t,t') C^{\mp \pm}(q,t,t') 
			\nonumber \\
& & \hspace{1cm} + T_{14}(k,p,q) G^{\pm \pm}(k,t-t') 
		   C^{\pm \pm}(p,t,t') C^{\mp \mp}(q,t,t') 
			\nonumber \\
& & \hspace{1cm} + T_{15}(k,p,q) G^{\pm \mp}(k,t-t') 
                   C^{\pm \pm}(p,t,t') C^{\mp \mp}(q,t,t') 
			\nonumber \\
& & \hspace{1cm} + T_{16}(k,p,q) G^{\pm \mp}(k,t-t') 
                   C^{\pm \mp}(p,t,t') C^{\mp \pm}(q,t,t') 
			\nonumber \\
& &\hspace{1cm}  + T_{17}(k,p,q) G^{\pm \pm}(p,t-t') 
                   C^{\pm \mp}(k,t,t') C^{\mp \pm}(q,t,t') 
			\nonumber \\
& &\hspace{1cm}  + T_{18}(k,p,q) G^{\pm \mp}(p,t-t') 
                   C^{\pm \pm}(k,t,t') C^{\mp \mp}(q,t,t')
			\nonumber \\
& &\hspace{1cm}  + T_{19}(k,p,q) G^{\pm \mp}(p,t-t') 
                   C^{\pm \pm}(k,t,t') C^{\mp \mp}(q,t,t')
			\nonumber \\
& &\hspace{1cm}  + T_{20}(k,p,q) G^{\pm \mp}(p,t-t') 
                   C^{\pm \pm}(k,t,t') C^{\mp \mp}(q,t,t')
			\nonumber \\
& &\hspace{1cm}  + T_{21}(k,p,q) G^{\mp \pm}(q,t-t') 
                   C^{\pm \mp}(k,t,t') C^{\pm \pm}(p,t,t')
			\nonumber \\
& &\hspace{1cm}  + T_{22}(k,p,q) G^{\mp \pm}(q,t-t') 
                   C^{\pm \pm}(k,t,t') C^{\pm \mp}(p,t,t')
			\nonumber \\
& &\hspace{1cm}  + T_{23}(k,p,q) G^{\mp \mp}(q,t-t') 
                   C^{\pm \pm}(k,t,t') C^{\pm \mp}(p,t,t')
			\nonumber \\
& &\hspace{1cm}  + T_{24}(k,p,q) G^{\mp \mp}(q,t-t') 
                   C^{\pm \mp}(k,t,t') C^{\pm \pm}(p,t,t')]
\eea
where $T_i(k,p,q)$ are given in Appendix A.  

Now we use the approximation that $r$ is small. 
In terms of renormalized
$\hat{\nu}$ matrix
\bea
\hat{\nu}(k) = 
\left(
\begin{array}{cc}
r \zeta & \alpha \\ 
r \psi  & \beta
\end{array} 
\right),
\eea
the Green's function  
$\hat{G}(k,t-t') = \exp{- [\hat{\nu} k^2 (t-t')]}$ to leading
order in $r$ is 
\be
\hat{G}(k,t-t') = 
\left(
\begin{array}{cc}
1-\frac{r \alpha \psi}{\beta^2} (1-e^{-\beta(t-t')}) &
- \left\{ \frac{\alpha}{\beta}+\frac{r \alpha}{\beta} 
         \left( \frac{\zeta}{\beta}- \frac{2 \alpha \psi}{\beta^2} \right)
  \right\}  (1-e^{-\beta (t-t')}) \\
-\frac{r \psi}{\beta} (1-e^{-\beta (t-t')}) &
e^{-\beta (t-t')} +\frac{r \alpha \psi}{\beta^2} 
			(1-e^{-\beta (t-t')}) 
\end{array}
\right) .
\ee
For derivation and further details on the renormalized $\hat{\nu}$, 
refer to Verma \cite{MKV:MHDRG}.
The correlation matrix $\hat{C}(k,t-t')$ is given by
\bea
\left(
\begin{array}{cc}
C^{++}(k,t,t') & C^{+-}(k,t,t') \\ 
C^{-+}(k,t,t') & C^{--}(k,t,t') 
\end{array}
\right) 
& = & 
\hat{G}(k,t-t')
\left(
\begin{array}{cc}
C^{++}(k) & C^{+-}(k) \\ 
C^{-+}(k) & C^{--}(k) 
\end{array}
\right) 
\eea

The quantities $C^{\pm \pm}(k)$ can be written in terms of
$E^\pm(k)$ as
\be
C^{\pm \pm}(k) = \frac{2 (2 \pi)^d}{S_d (d-1)} k^{-(d-1)} E^{\pm}(k).
\ee
We take $C^{\pm \mp}(k)=0$ (or $r_A=1$) for simplifying
the calculation.
We take Kolmogorov's spectrum for $E^{\pm}(k)$ [see Eq.~(\ref{eqn:KolmMHD})],
and
\be
\hat{\nu}(k) = \left(
\begin{array}{cc}
r \zeta & \alpha \\ 
r \psi  & \beta
\end{array}
\right) =
\left(
\begin{array}{cc}
r \zeta^* & \alpha^* \\ 
r \psi^*  & \beta^*
\end{array}
\right)  \sqrt{K^+}
\frac{\left( \Pi^{+} \right)^{2/3}} {\left( \Pi^{-}
	\right)^{1/3}} k^{-4/3}
\ee
The renormalized parameters $\zeta^*, \alpha^*, \psi^*$, and $\beta^*$
have been calculated in Verma \cite{MKV:MHDRG}. Finally, the matrices
$\hat{G}(k,t-t')$ and $\hat{C}(k,t,t')$ can be written in in terms of
renormalized parameters and Kolmogorov's spectrum.  

Now we substitute $\hat{G}(k,t-t')$ and $\hat{C}(k,t,t')$ in
Eq.~(\ref{eqn:Spm}), and keep terms only to the leading order in $r$.
We find that the terms (1,4,5,8,9,10) of Eq.~(\ref{eqn:Spm}) vanish.
The final equation for the fluxes $\Pi^{\pm}$ to the leading order in
$r$ are
\vspace{0.3cm}
\be
\Pi^{\pm} =  r  \frac{(\Pi^+)^2}{\Pi^-}  (K^+)^{3/2} 
	\left[ \frac{4 S_{d-1}}{(d-1)^2 S_d} \int_0^1 dv \ln{(1/v)}
		\int_{1-v}^{1+v} dw 
               	(vw)^{d-2} (\sin \alpha)^{d-3} F^{\pm}(v,w) 
        \right]
	\label{eqn:Pipm}
\ee
\vspace{0.3cm}
where the integrand $F^{\pm}$ are 
\bea 
F^+ & = &  t_{13}(v,w) (vw)^{-d-2/3}
		\frac{1}{\beta^* w^{2/3}} 
          +t_{14}(v,w) (vw)^{-d-2/3} \frac{\alpha^*}{\beta^*} 
             \left\{ \frac{1}{\beta^*(1+w^{2/3})}
               -\frac{1}{\beta^* w^{2/3}} \right\} \nonumber \\
& & 	  +t_{15}(v,w) w^{-d-2/3} \frac{1}{\beta^* w^{2/3}}
          +t_{16}(v,w) w^{-d-2/3} \frac{\alpha^*}{\beta^*} 
             \left\{ \frac{1}{\beta^*(v^{2/3}+w^{2/3})}
               -\frac{1}{\beta^* w^{2/3}} \right\} \nonumber \\
& &	  +t_{17}(v,w) v^{-d-2/3} \frac{\alpha^*}{\beta^*} 
             \left\{ \frac{1}{\beta^*(v^{2/3}+w^{2/3})}
               -\frac{1}{\beta^* w^{2/3}} \right\} \nonumber \\
& &	  +t_{18}(v,w) v^{-d-2/3} \frac{\alpha^*}{\beta^*} 
             \left\{ \frac{1}{\beta^*(1+w^{2/3})}
               -\frac{1}{\beta^* w^{2/3}} \right\} 	\\
F^- & = & t_{13}(v,w) (vw)^{-d-2/3}
		\frac{1}{\beta^* (1+v^{2/3})}
         +t_{15}(v,w) w^{-d-2/3}
             	\frac{1}{\beta^* (1+v^{2/3})} 
\eea
We denote the bracketed term of Eq.~(\ref{eqn:Pipm}) by $I^{\pm}$
and compute them numerically.  We find that the integrals are finite for
$d=2$ and 3.  Also note that $I^{\pm}$ are independent of $r$. We
calculate the constant $K^\pm$ of Eq.~(\ref{eqn:Pipm}) in terms of
$I^{\pm}$; the constants $K^\pm$ are listed in Table~\ref{tab:Kpm} for
various values of $r$ in $d=2$ and 3.  The constants $K^{\pm}$ depend
very sensitively on $r$.  Also, there is a change of behaviour near
$r=(I^-/I^+)^2=r_c$; $K^- < K^+$ for $r<r_c$, whereas inequality
reverses for $r$ beyond $r_c$.

From the equations derived above, we can derive many important
relationships. For example,
\begin{equation}
\frac{\Pi^-}{\Pi^+} = \frac{I^-}{I^+}
\end{equation}
Since $I^{\pm}$ are independent of $r$, we can immediately conclude that
the ratio $\Pi^-/\Pi^+$ is also {\em independent of $r$}. 
This is an important conclusion from our calculation. 
From the above equations we can also
derive 
\begin{eqnarray}
K^{+} & = & \frac{1}{r^{2/3}} \frac{(I^-)^{2/3}}{(I^+)^{4/3}}  \\
K^{-} & = & r^{1/3} \frac{(I^+)^{2/3}}{(I^-)^{4/3}}  \\
\frac{K^-}{K^+} & = & r \left( \frac{I^+}{I^-} \right)^2
\end{eqnarray}

The total energy cascade rate can  be written in terms of $E^+(k)$ as
\begin{equation}
\Pi = \frac{1}{2}(\Pi^+ + \Pi^-) =
     \frac{r}{2} (I^+ + I^-) (E^+(k))^{3/2} k^{5/2}
\label{eqn:flux_sigc1}
\end{equation}
Since $I^{\pm}$ is independent of $r$, $\Pi$ is a linear function of $r$. 
When we apply the above formula to the solar wind stream with
$r=0.07$, we find that $K^+=2.12$ and $K^-=0.85$. 

As mentioned in the previous section, the observed temperature
evolution was studied by Verma {\em et al.}~\cite{MKV:SWheat} and Tu \cite{Tu}.
For streams with $\sigma_c \rightarrow 1$, Verma {\em et al.}~\cite{MKV:SWheat}
had  assumed that $K^{+}=K^{-}=K$ independent of $\sigma_c$,  and 
derived the total turbulent dissipation rate to be
\be
\label{SW}
\Pi = \frac{r+\sqrt{r}}{2 K^{3/2}} (E^+(k))^{3/2} k^{5/2}.
\label{eqn:MKVSWheat}
\ee 
Clearly the assumption that $K^{+}=K^{-}$, as well as the above
formula~(\ref{eqn:MKVSWheat}) is incorrect.  Hence, the calculation of
Verma {\em et al.}~\cite{MKV:SWheat} needs to be modified.  
The substitution of the parameters $r$ and $K^{\pm}$ in our
formula~(\ref{eqn:flux_sigc1}) gives us an estimate of the turbulent
heating that is an order of magnitude higher than the observed overall
heating in the solar wind \cite{MKV:SWheat}.   Some of the resolutions of
this paradox are: (1) the assumption that the solar wind has
reached steady-state is incorrect, and the
formula~(\ref{eqn:flux_sigc1}) is inapplicable to the solar wind
streams with large $\sigma_c$; or (2) the constants $K^{\pm}$
calculated above will be modified significantly by the mean magnetic
field, anisotropy, and helicity etc.  In case of the former, one
needs to understand the nonequilibrium evolution of MHD turbulence,
while in case of the latter, the field theoretic calculation has
to generalized in presence of mean magnetic field and helicity.  Both
these generalizations are beyond the scope of this paper.

\section{Summary and Conclusions}

In this paper we have theoretically calculated various energy cascade
rates in the inertial range of  {\em nonhelical} MHD turbulence.
Our procedure is based on field-theoretic approach.  Using the
steady-state condition we also calculate the energy supply rate from the
large-scale velocity field to the large-scale magnetic field.  For
simplicity of the calculation, we have taken two special cases: (1)
$\sigma_c=0$; (2) $\sigma_c \rightarrow 1$.  Throughout the
calculation we assume that the velocity modes at large length-scales
are forced.

We will first summarize the results for $\sigma_c=0$ case in $d=3$.
The cascade rates \piulbg, \piblbg, \piulbl, \piblug are approximately
the same for $r_A$ in the range of $0.5-1$, but the flux \piulug is
rather small.  The sign of \piblbg is positive, indicating that the ME
cascades forward, that is from large length-scales to small
length-scales.  The large-scale magnetic field is maintained by the
$\Pi^{u<}_{b<}$ flux.   We exploit this result to construct
a dynamo model for galaxy.  This result is discussed in paper II.

Recently Cho and Vishniac (CV) \cite{ChoVish:gene} performed numerical
simulation of nonhelical MHD turbulence and arrived at the following
conclusion based on their numerical results.  In our language, their
results for large $r_A$ can be rephrased as (1) $\Pi^{u<}_{u>} \approx
U^3$; (2) $\Pi^{u<}_{(b<+b>)} \approx U B^2$; (3) $\Pi^{u<}_{b<}
\approx (U- cB) B^2$, where $U$ and $B$ are the large-scale velocity
and magnetic field respectively, and $c$ is a constant.  When we
compare our theoretical findings with CV's result, we find our results
can explain CV's first and second results, but they are only partly
consistent with the third result.  From Eq.~(\ref{eqn:Fuu}) it can be
easily seen that \piulug depends on the KE in the same manner as in
fluid turbulence.  Hence, $\Pi^{u<}_{u>} \approx U^3/L$, a result
consistent with the first result of CV.  Using $\Pi^{u<}_{(b<+b>)} =
\Pi^{u<}_{b>}+\Pi^{b<}_{b>}+ \Pi^{b<}_{u>}$ and the definitions of
$F's$ [Eqs.~(\ref{eqn:Fbu}-\ref{eqn:Fbb})] we can easily show that
\be
\frac{\Pi^{u<}_{(b<+b>)}}{\Pi} \approx ... r_A^{-1} + ... r_A^{-2} 
\ee
where $...$ represents a constant.
We estimate the above equation in the large $r_A$ limit ($E^u \gg
E^b$).  In this limit, $\Pi \approx U^3/L$.  Hence, to a leading order
in $r_A^{-1}$ 
\be 
\Pi^{u<}_{(b<+b>)} \approx \Pi \frac{E^b}{E^u} \approx U B^2/L 
\ee 
From Eqs.~(\ref{eqn:Fuu}-\ref{eqn:Fbb}), we also conclude that 
\be 
\Pi^{u<}_{b<} \approx (... r_A^{-1} + ... r_A^{-2}) \Pi \approx 
\left(... \frac{U}{L} - ... \frac{B^2}{U L} \right) B^2
\ee 

Note that the first part of the above equation matches with CV's
first part, but the second part of $\Pi^{u<}_{b<}$ differs from CV's
result by a factor of $B/U$.  Since $B/U \approx 1$ at steady state,
it is difficult to differentiate our results with that of CV.  On the
whole, our theoretical calculation is able to explain the numerical
results of CV.

For $d=3$ the Kolmogorov's constant $K$ is approximately constant and
is close to 1.5 for all $r_A$ greater that 1/2, same as that for fluid
turbulence ($r_A=\infty$).  This result implies that the total cascade
rate does not change appreciably under the variation of $r_A$ (since
$\Pi \propto K^{-3/2}$).  The cascade rates vanish near $r_A = 0.25$;
this result is in the expected lines because MHD equations become
linear in $r_A =0$ limit.  Comparison with the past results shows that
our result differs from that of Verma and Bhattacharjee's calculation
\cite{MKVJKB} where Kolmogorov's constant changes significantly with
the variation of $r_A$.  Note, however, that our procedure described
here is an improvement over that of Verma and Bhattacharjee, where
they had assumed a wavenumber cutoff for the self-energy integral for
curing the infrared divergence problem.  They had also assumed a
specific type of self-energy matrix which can be shown to be correct
only in some regime.

When we vary $d$, we find that for large $d$,
$\Pi^{u<}_{u>}=\Pi^{u<}_{b>}= \Pi^{b<}_{b>}=\Pi^{b<}_{u>}$.  In
addition we also observe that Kolmogorov's constant MHD turbulence
increases with dimensions as $d^{1/3}$.  The same variation is
observed for fluid turbulence \cite{FourFris}.  This result indicates
that the cascade rates decrease in higher dimensions.  We could
calculate fluxes for $d \ge 2.2$ because the RG fixed point is
unstable for dimensions lower than 2.2 (Verma \cite{MKV:MHDRG}).  However,
the RG fixed point for fluid turbulence is stable for $d =2$, and the
Kolmogorov's constant in the inverse cascade regime of 2D fluid
turbulence comes out to be 6.3.  For this computation, the
renormalized viscosity $\nu^*$ was taken as -0.6.  It is interesting
to note that Dar {\em et al.}~\cite{Dar:flux} find negative KE flux
(\piulug) in their 2D MHD turbulence simulation; this is reminiscent of
2D fluid turbulence.

In the other extreme limit $\sigma_c \rightarrow 1$ and $r_A=1$, we
find that Kolmogorov's constants $K^{+}$ and $K^-$ are not equal, and
the ratio $K^-/K^+$ depends very sensitively on $r=E^{-}(k)/E^{+}(k)$.
Both the fluxes $\Pi^\pm$, and also the total flux $\Pi$, are
proportional to $r$.  The flux ratio $\Pi^-/\Pi^+$ is found to be
independent of $r$.  We also discuss the implications of our flux
results to the heating of the solar wind.

In this paper we have restricted ourselves to nonhelical turbulence.
Helical MHD turbulence is very important specially for the growth of
magnetic energy (dynamo).  The energy fluxes for helical MHD have been
discussed in paper II.  The study of the effects of mean
magnetic field using field theory has been relegated for future.

\acknowledgements The author thanks J. K. Bhattacharjee, Gaurav Dar,
and V. Eswaran for discussion and suggestions.  He also thanks
Mustansir Barma (TIFR, Mumbai) and Krishna Kumar (ISI, Calcutta) for
useful suggestions and kind hospitality during his stay in their
institutes on his sabbatical leave.

\appendix
\section{Values of $T_i$}

The algebraic expressions for $T_i(k,p,q)$ are given below.
\begin{eqnarray}
T_1(k,p,q) & = & k_i P^{+}_{jab}(k) P_{ja}(p) P_{ib}(q)
			 \nonumber \\
           & = & kp\left( (d-3)z + (d-2)xy +2 z^3+ 2 x y z^2
			  + x^2 z \right) \\
T_3(k,p,q) & = & k_i P^{-}_{jab}(k) P_{ja}(p) P_{ib}(q)
			 \nonumber \\
           & = & -k^2 \left( (d-2)(1-y^2) + z^2 +x y z \right) \\
T_5(k,p,q) & = & -k_i P^{+}_{jab}(p) P_{ja}(k) P_{ib}(q)
			\nonumber \\
           & = & -kp\left( (d-3)z + (d-2)xy +2 z^3+ 2 x y z^2
			  + y^2 z \right) \\
T_7(k,p,q) & = & -k_i P^{-}_{jab}(p) P_{ja}(k) P_{ib}(q)
			\nonumber \\
           & = & -kp\left( (2-d)x y + (1-d)z +y^2 z \right) \\
T_9(k,p,q) & = & -k_i P^{+}_{iab}(q) P_{ja}(k) P_{jb}(p)
			\nonumber \\
           & = & -k q \left(x z - 2 x y^2 z - y z^2 \right) \\
T_{11}(k,p,q) & = & -k_i P^{-}_{iab}(q) P_{ja}(k) P_{jb}(p)
			\nonumber \\
           & = & -k q z \left(x + y z \right) \\
T_{2 n}(k,p,q) & = & -T_{2 n -1 }(k,p,q) 
			\hspace{1cm} \mbox{for $n=1..6$}\\
T_{13,15}(k,p,q) & = & k_i M_{jab}(k') P_{ja}(p) P_{ib}(q)
			 \nonumber \\
           & = & - k p  y z (y+x z)  \\
T_{14,16}(k,p,q) & = & k_i M_{jab}(k') P_{jb}(p) P_{ia}(q)
			 \nonumber \\
           & = & k^2 (1-y^2) (d-2+z^2) \\
T_{17,19}(k,p,q) & = & k_i M_{jab}(p) P_{ja}(k) P_{ib}(q) \nonumber \\
	      & = & k p  x z (x+y z)  \\
T_{18,20}(k,p,q) & = & -T_{14}(k,p,q) \\
T_{21,23}(k,p,q) & = & k_i M_{iab}(q) P_{ja}(k) P_{jb}(p) \nonumber \\
	      & = & -k p x y (1-z^2) \\
T_{22,24}(k,p,q) & = & -T_{13}(k,p,q) 
\end{eqnarray}
where ${\bf k=p+q}$, and  $x,y,z$ are defined by
\be
{\bf p \cdot q} = -pqx; \hspace{1cm} {\bf q \cdot k}= qky; \hspace{1cm} {\bf p \cdot k}=pkz.
\ee


\vspace{1in}
\centerline{\large Figure Captions}
\vspace{1cm}

\noindent
{\bf Fig. 1.} \, Various energy cascade rates of MHD turbulence.  The
illustrated wavenumber spheres contain ${\bf u<}$ and ${\bf b<}$
modes, while ${\bf u>}$ and ${\bf b>}$ are modes outside these
spheres.  The velocity fields is forced at large-scale.

\newpage

\begin{table}
\caption{The flux ratios computed by Dar {\em et al.}
($d=2, \sigma_c \approx 0, r_A \approx 0.5$)}
\label{tab:flux_Dar}
\begin{ruledtabular} 
\begin{tabular}{lccccr}
 \piulug/$\Pi$ & \piulbg/$\Pi$ & \piblug/$\Pi$ & \piblbg/$\Pi$ 
		& \piulbl/$\Pi$ & $K^{+}$ \\ \hline
$-0.13$ & 0.68 & $-0.09$ & 0.47 & 0.37 & $\approx 4$ \\
\end{tabular}
\end{ruledtabular} 
\end{table}
\begin{table}
\caption{The computed values of energy cascade rates of MHD 
turbulence  for various $r_A$  when ${\bf d=3}$
and $\sigma_c=0$.} 
\label{tab:flux_d3}
\begin{ruledtabular}  
\begin{tabular}{lccccccl} 
$\Pi \setminus r_A$  & 
             5000        & 100  & 5    & 1    & 0.5   & 0.3  & Trend \\ \hline
								 
\piulug/$\Pi$ & 1        & 0.97 &0.60  & 0.12 & 0.037 & 0.011& decreases \\
								
\piulbg/$\Pi$ & 3.5E-4  &1.7E-2& 0.25& 0.40 & 0.33  & 0.36 & increases \\
              &         &      &      &      &       &      & then saturates \\
							
\piblug/$\Pi$ &-1.1E-4  &-5.1E-3&$-0.05$&0.12& 0.33  & 0.42 & increases \\
              &         &      &      &      &       &      & then saturates \\
								
\piblbg/$\Pi$ & 2.7E-4  &1.3E-2& 0.20& 0.35 & 0.30  & 0.21 & increases \\
              &         &      &     &      &       &      & then dips \\
							
\piulbl/$\Pi$ & 1.7E-4  &8.1E-3& 0.15 & 0.47 & 0.63  & 0.63 & increases \\
              &         &      &      &      &       &      & then saturates \\
								
$K^+$         & 1.53    & 1.51 & 1.55 & 1.50 & 1.65  & 3.26 & approx. const \\
              &         &      &      &      &       &      &  till $r_A 
							\approx 0.5$ \\
							
$K^u$         & 1.53    & 1.50 & 1.29 & 0.75 & 0.55  & 0.75 & decreases \\
\end{tabular}
\end{ruledtabular} 
\end{table}
\begin{table}
\caption{The computed values of energy cascade rates of MHD 
turbulence  for various space dimensions $d$  when
$\sigma_c=0$ and $r_A=1$.}
\label{tab:flux_alldra1}
\begin{ruledtabular} 
\begin{tabular}{lccccccr} 
$\Pi \setminus d$  & 
               2.1& 2.2  & 2.5   & 3    & 4     & 10   & 100  \\ \hline

\piulug/$\Pi$ & - & 0.02 & 0.068& 0.12  & 0.17  & 0.23 & 0.25 \\
\piulbg/$\Pi$ & - & 0.61 & 0.49 & 0.40  & 0.34  & 0.27 & 0.25 \\
\piblug/$\Pi$ & - &$-0.027$ & 0.048& 0.12& 0.18 & 0.23 & 0.25 \\
\piblbg/$\Pi$ & - & 0.40 & 0.39 & 0.35  & 0.31  & 0.27 & 0.25 \\
\piulbl/$\Pi$ & - & 0.37 & 0.4 & 0.47  & 0.49  & 0.50 & 0.50 \\ \hline
$K^+$         & - & 1.4 & 1.4 & 1.50  & 1.57  & 1.90 & 3.46 \\ 
$K^u$         & - & 0.69 & 0.72 & 0.75  & 0.79  & 0.95 & 1.73  \\
\end{tabular}
\end{ruledtabular}
\end{table}

\begin{table}
\caption{The computed values Kolmogorov's constants
for $\sigma_c \rightarrow 1$ and $r_A =1$ limit for various $r=E^-/E^+$
($d=2,3$) }
\label{tab:Kpm}
\begin{ruledtabular} 
\begin{tabular}{lccr}
 $d$ & $r$      & $K^+$ & $K^{-}$  \\ \hline
      & 0.17     & 1.4  & 1.4     \\
      & 0.10     & 2.1  & 1.2     \\
  3   & 0.07     & 2.7  & 1.07     \\
      &$10^{-3}$ & 45    & 0.26     \\
      &$10^{-6}$ & 4528  & 0.026    \\ \hline
      
      & 0.1      & 1.2  & 2.4   \\
      & 0.07     & 1.5  & 2.2   \\
  2   &0.047     & 1.9  & 1.9   \\
      &$10^{-3}$ & 25    & 0.52  \\
      &$10^{-6}$ & 2480  & 0.052 \\
\end{tabular}
\end{ruledtabular}
\end{table}

\begin{figure}
\centerline{\psfig{figure=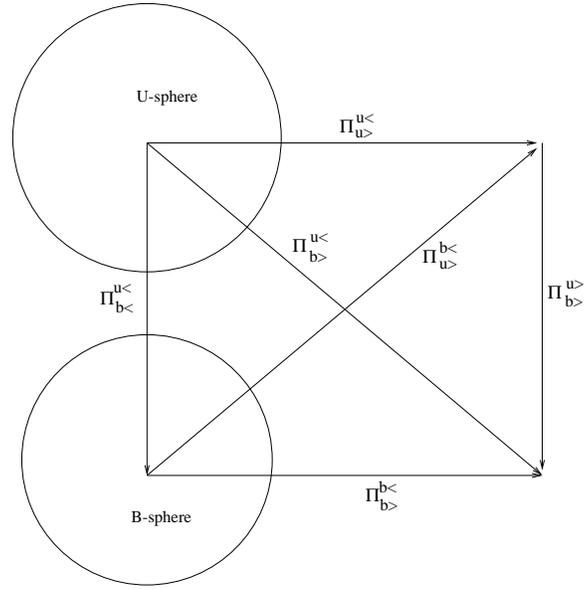,width=3in,angle=0}}
        \vspace*{0.5cm}
\caption{Various energy cascade rates of MHD turbulence.  The
illustrated wavenumber spheres contain ${\bf u<}$ and ${\bf b<}$
modes, while ${\bf u>}$ and ${\bf b>}$ are modes outside these
spheres.  The velocity fields is forced at large-scale. }
\label{fig:mhdflux}
\end{figure}
\end{document}